\documentclass[apj]{emulateapj}
\submitted{Accepted for publication by The Astrophysical Journal Letters on 2012, January 3. }

\usepackage{color}
\usepackage{natbib}

\newcommand\ionB[2]{#1$\;${\scshape{#2}}}

\def\ForbSII{\mbox{[\ionB{S}{ii}]}}

\def\Ha{\ifmmode^{\mathrm{H}\alpha }\else$\mathrm{H}\alpha$\fi}
\def\Hb{\ifmmode^{\mathrm{H}\beta }\else$\mathrm{H}\beta$\fi}
\def\LyA{\ifmmode^{\mathrm{H}\alpha }\else$\mathrm{Ly}\alpha$\fi}
\def\BrA{\ifmmode^{\mathrm{Br}\alpha }\else$\mathrm{Br}\alpha$\fi}
\def\BrG{\ifmmode^{\mathrm{Br}\gamma }\else$\mathrm{Br}\gamma$\fi}
\def\PaB{\ifmmode^{\mathrm{Pa}\beta }\else$\mathrm{Pa}\beta$\fi}
\def\mag{\ifmmode^{\rm m }\else$^{\rm m}$\fi}

\def\as{$\,^{\prime\prime}\,$}

\def\hh{\ifmmode^{\rm h}\else$^{\rm h}$\fi}
\def\mm{\ifmmode^{\rm m}\else$^{\rm m}$\fi}
\def\ss{\ifmmode^{\rm s}\else$^{\rm s}$\fi}
\def\deg{\ifmmode^\circ\else$^\circ $\fi}
\def\amin{\ifmmode^\prime\else$^\prime $\fi}

\def\decdm#1#2{\ifmmode{#1}\else{$#1$}\fi\deg\ #2\amin\ }

\def\dec#1#2#3{\ifmmode{#1}\else{$#1$}\fi\deg\ #2\amin\ #3\as\ }

\def\decb#1#2#3#4{\ifmmode{#1}\else{$#1$}\fi\deg\ #2\amin\ #3\farcs#4 }

\shorttitle{V921\,Scorpii: Discovery of a close companion and relation to the large-scale bipolar nebula}
\shortauthors{Kraus et al.}

\begin{document}




\title{On the nature of the Herbig~B[e] star binary system V921\,Scorpii:\\
  Discovery of a close companion and relation to the large-scale bipolar nebula\footnotemark[1]}


\footnotetext[1]{Based on observations made with ESO telescopes 
at the Paranal Observatory under the open-time programme ID
081.C-0706(A-D) and with the Magellan Baade and Clay telescopes.
}


\author{Stefan Kraus\altaffilmark{1},
  Nuria Calvet\altaffilmark{1},
  Lee Hartmann\altaffilmark{1},
  Karl-Heinz Hofmann\altaffilmark{2},
  Alexander Kreplin\altaffilmark{2},
  John D.\ Monnier\altaffilmark{1}, and
  Gerd Weigelt\altaffilmark{2}}


\affil{
$^{1}$~Department of Astronomy, University of Michigan, 918 Dennison Building, Ann Arbor, MI 48109-1090, USA\\
$^{2}$~Max Planck Institut f\"ur Radioastronomie, Auf dem H\"ugel 69, 53121 Bonn, Germany
}


\begin{abstract}
Belonging to the group of B[e] stars, V921\,Scorpii is associated with a 
strong infrared excess and permitted and forbidden line emission, indicating the presence of 
low- and high-density circumstellar gas and dust.
Many aspects of V921\,Sco and other B[e] stars still remain mysterious, 
including their evolutionary state and the physical conditions resulting 
in the class-defining characteristics.
In this Letter, we employ VLTI/AMBER spectro-interferometry in order to 
reconstruct high-resolution ({$\lambda/2B=0.0013$\arcsec}) 
model-independent interferometric images for three wavelength bands around 1.65, 2.0, and 2.3~$\mu$m.
In our images, we discover a close ($25.0 \pm 0.8$~milliarcsecond, corresponding to $\sim 29 \pm 0.9$~AU at 1.15~kpc) companion around V921\,Sco.
Between two epochs in 2008 and 2009, we measure orbital motion of $\sim 7^{\circ}$,
implying an orbital period of $\sim 35$~years (for a circular orbit).
Around the primary star, we detect a disk-like structure 
with indications for a radial temperature gradient.
The polar axis of this AU-scale disk is aligned with the arcminute-scale bipolar nebula 
in which V921\,Sco is embedded.  Using Magellan/IMACS imaging, we detect multi-layered 
arc-shaped sub-structure in the nebula, suggesting episodic outflow activity from the system 
with a period of $\sim 25$~years, roughly matching the estimated orbital period of the companion.
Our study supports the hypothesis that the B[e] phenomenon is related to dynamical 
interaction in a close binary system.
\end{abstract}


\keywords{stars: pre-main sequence ---
  stars: individual (V921\,Scorpii) --- 
  binaries: close ---
  protoplanetary disks --- 
  accretion, accretion disks --- 
  techniques: interferometric}

\section{Introduction}
B[e] stars are intermediate-mass stars associated with substantial amounts
of circumstellar gas and dust, as indicated by permitted and forbidden line emission,
in particular of [\ion{O}{1}] and [\ion{Fe}{2}], and a strong infrared excess \citep{all76}.
These class-defining characteristics have been observed 
in a wide range of evolutionary stages \citep{lam98},
including pre-main-sequence stars (Herbig~Ae/B[e]),
post-main-sequence stars (supergiants, symbiotic stars, or compact planetary nebulae),
and stars of unknown nature (unclassified B[e] stars).
In the pre-main-sequence stage, about half of all 
intermediate-mass young stars show the 
{\it B[e] phenomenon} \citep{oud06}, although significant differences in the
strength of the forbidden line emission can be observed.

It has been proposed that the B[e] phenomenon might be related 
to the presence of a close binary system, where the line-emitting gas is ejected in
recurring mass-loss events triggered by the companion \citep[e.g.][]{she00,mir07}.
To test this hypothesis, it is important to search for close companions around B[e] stars.
Given the kiloparsec distance of B[e] stars, this task requires 
high angular resolution, which we achieve in this Letter using infrared interferometry.

We observed the unclassified B[e] star V921\,Sco using the Very Large Telescope Interferometer (VLTI).
The AU-scale environment around this B0Ve-type star has been investigated by two earlier
interferometric studies, which determined the spatial extension of the 
Br$\gamma$-line emitting region \citep{kra08b}
and the structure of the continuum-emitting disk \citep{kre12}.
V921\,Sco is associated with an intriguing nebula, which shows 
reflection as well as absorption characteristics \citep{hut90}
and which has been imaged both at visual \citep{hut90} and mid-infrared wavelengths \citep{boe09}.
There is some debate on the distance \citep[$\sim 1.15$~kpc,][]{bor07} and 
nature of V921\,Sco, with authors arguing both for an
evolved \citep{hut90,bor07} and
young \citep{ben98,hab03,ack05,ack06}
evolutionary stage.

In this Letter, we report on Magellan wide-field imaging 
and VLTI aperture synthesis imaging observations (Sect.~\ref{sec:observations}), 
which reveal the presence of a close companion around V921\,Sco.
The astrometry of the binary system and the orientation of the circumprimary disk
will be derived using quantitative modeling (Sect.~\ref{sec:continuum}) and
interpreted in Sect.~\ref{sec:interp}.
A summary of our results will be presented in Sect.~\ref{sec:conclusions}.

\section{Observations}
\label{sec:observations}

\subsection{VLTI/AMBER spectro-interferometry}
\label{sec:obsAMBER}

Our observations on V921\,Sco were carried out using 
the AMBER instrument \citep{pet07},
which allowed us to coherently combine the light from 
three of the VLTI 1.8\,m auxiliary telescopes.
The observations were conducted using AMBER's
low spectral resolution mode, which covers
the near-infrared $H$- and $K$-bands
($1.44$ to 2.50~$\mu$m) with a spectral resolution $R \sim 35$.
The observations were carried out between
2008-04-03 and 2009-03-19
on four different array configurations (Tab.~\ref{tab:obslog}),
providing a good $uv$-coverage
with baseline lengths between 10 and 127\,m.
For all observations, we used a detector integration time (DIT) of 100\,ms.

For a significant fraction of our data we find that the 
closure phases (CPs) vary significantly between subsequent exposures 
(i.e.\ on time scales of several minutes).  As discussed in Sect.~\ref{sec:continuum},
these rapid variations are very likely due to the presence of a companion star.
Therefore, in contrast to the
standard AMBER data reduction procedure, we decided not to average the
individual exposures, but to fit the quantities derived 
from the individual data exposures separately.

Each observation on V921\,Sco was accompanied by
observations on interferometric calibrator stars of known 
intrinsic diameter,
allowing us to monitor the instrumental and atmospheric transfer function.
Both for the science and calibrator star observations, 
we extract raw visibilities and CPs
using the amdlib (V3.0) data reduction software \citep{tat07b,che09}.
In order to associate the CP sign with the
on-sky orientation, we use a reference 
data set\footnote{The reference data set can
be accessed on the website http://www.skraus.eu/files/amber.htm}
on the binary star $\theta^1$\,Orionis~C \citep{kra09a}.

\begin{deluxetable}{ccccc}
\tabletypesize{\scriptsize}
\tablecolumns{5}
\tablewidth{0pc}
\tablecaption{Observation log of our VLTI/AMBER observations \label{tab:obslog}}
\tablehead{
  \colhead{Date}        & \colhead{Time}  & \colhead{NEXP}  & \colhead{Telescope}   &  \colhead{Calibrator} \\
  \colhead{(UT)}        & \colhead{(UT)}  &                 & \colhead{triplet}     &            
}
\startdata
  2008-05-21 & 09:20  & 10    & A0-D0-H0    & HD\,159941  \\
  2008-05-22 & 04:18  & 9     & A0-D0-H0    & HD\,159941  \\
  2008-05-22 & 05:25  & 9     & A0-D0-H0    & HD\,159941  \\
  2008-05-22 & 06:47  & 9     & A0-D0-H0    & HD\,159941  \\
  2008-05-24 & 06:29  & 11    & A0-D0-H0    & HD\,159941  \\
  2008-05-24 & 09:24  & 9     & A0-D0-H0    & HD\,159941  \\
  2008-09-21 & 01:54  & 5     & A0-K0-G1    & HD\,159941  \\
  2008-04-28 & 08:32  & 5     & D0-H0-G1    & HD\,163197  \\
  2008-04-28 & 09:18  & 5     & D0-H0-G1    & HD\,163197  \\
  2008-05-26 & 07:28  & 5     & D0-H0-G1    & HD\,159941  \\
  2008-07-04 & 02:59  & 5     & D0-H0-G1    & HD\,159941  \\
  2008-07-04 & 04:02  & 5     & D0-H0-G1    & HD\,159941  \\
  2008-07-04 & 04:46  & 5     & D0-H0-G1    & HD\,159941  \\
  2008-07-05 & 01:15  & 5     & D0-H0-G1    & HD\,159941  \\
  2008-07-05 & 06:24  & 5     & D0-H0-G1    & HD\,159941  \\
  2008-04-03 & 05:52  & 8     & E0-G0-H0    & HD\,152040  \\
  2008-04-05 & 07:16  & 5     & E0-G0-H0    & HD\,159941  \\
  2008-04-05 & 08:08  & 12    & E0-G0-H0    & HD\,159941  \\
  2008-06-04 & 07:03  & 5     & E0-G0-H0    & HD\,159941  \\
  2008-06-04 & 08:54  & 5     & E0-G0-H0    & HD\,159941  \\
  2008-06-07 & 08:32  & 5     & E0-G0-H0    & HD\,159941  \\
  \tableline
  \noalign{\smallskip}
  2009-02-18 & 09:05  & 5     & D0-H0-G1    & HD\,159941  \\
  2009-02-19 & 08:58  & 5     & D0-H0-G1    & HD\,159941  \\
  2009-03-19 & 08:54  & 15    & D0-H0-G1    & HD\,159941  
\enddata
\tablecomments{
  NEXP lists the number of recorded exposures
  during each visit on the source, 
  with each exposure consisting of 1000 individual interferograms.
  For the calibrators, we adopted the following uniform disk diameters:
  \object{HD\,106979}: $0.886\pm0.011$\,mas,
  \object{HD\,152040}: $0.581\pm0.04$\,mas (JMMC SearchCal); 
  \object{HD\,145921}: $0.957\pm0.013$\,mas, 
  \object{HD\,137730}: $0.984\pm0.014$\,mas \citep{mer06}; 
  \object{HD\,163197}: $0.96\pm0.01$\,mas, 
  \object{HD\,159941}: $1.09\pm0.02$\,mas \citep{ric05}.
}
\end{deluxetable}

\subsection{Magellan/IMACS+FIRE wide-field imaging}
\label{sec:obsIMACS}

\begin{figure*}
  \centering
  \includegraphics[angle=0,scale=0.85]{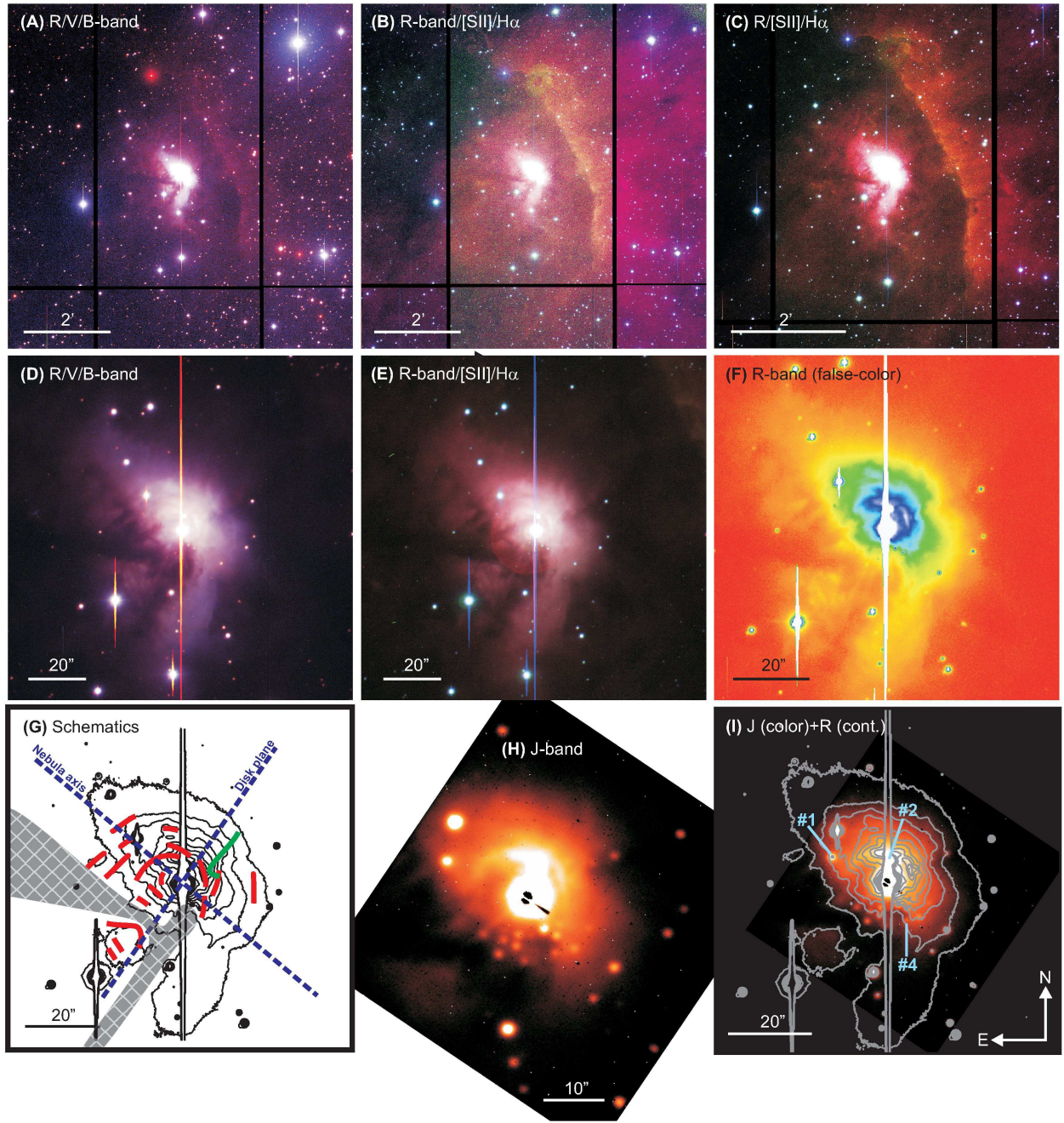}
  \caption{
    {\it (A), (D):} Magellan/IMACS composite images of V921\,Sco, taken in the filters
    Bessell-R (red), Bessell-V (green), and Bessell-B (blue) and shown for two different spatial scales.
    {\it (B), (C), (E):} IMACS images for Bessell-R (red), {\ForbSII} (green), and H$\alpha$ (blue).
    The black strips in panels {\it (A)} and {\it (B)} correspond to the gaps in the IMACS detector array.
    {\it (F):} IMACS image taken in Bessell-R filter, shown false-colors in order to emphasize some
    arc-like substructures.
    {\it (G):} Schematics, showing some of the filaments discovered discovered in our IMACS images,
    overplotted on Bessell-R contours.
    {\it (H):} $J$-band image recorded with the Magellan/FIRE acquisition camera
    and displayed with a heat color table and a square-root scaling.
    The FIRE spectrograph slit appears as a dark band roughly in the center of the image.
    {\it (I):} In the bottom right panel, we overplot the $J$-band image with the contours of the
    $R$-band image and also mark the location of the mid-infrared sources detected
    by \citet[source \#3 was not clearly detected in our $J$-band image,][]{hab03}.}
  \label{fig:IMACS}
\end{figure*}

In order to investigate the large-scale environment around V921\,Sco,
we employed the IMACS instrument on the Magellan/Baade 6.5\,m telescope.
The images were recorded on 2011-03-13 under exceptional atmospheric conditions (seeing FWHM $\sim$0.4\arcsec)
and cover a field of {$15.4$\arcmin} with a pixel size of 0.11\arcsec/pixel.
We employed the Bessell B, V, and R filters and two narrowband filters
centered around the {\ForbSII} and H$\alpha$ line (filters ``676circular'' and ``Halpha656'')
using DITs of 240\,s, 240\,s, 180\,s, 300\,s, and 360\,s, respectively.
The images were bias-subtracted and flat-fielded using standard IRAF data reduction routines
and are shown in Fig.~\ref{fig:IMACS}{\it A-F}.

Using the acquisition camera of the FIRE spectrograph \citep{sim08} at Magellan/Clay 
we also recorded on 2011-03-12 a $J$-band image of V921\,Sco (394$\times$394~pixels with a pixel size 0.147\arcsec/pixel),
which is shown in Fig.~\ref{fig:IMACS}{\it H}.

\section{Results: Continuum geometry}
\label{sec:continuum}

\subsection{Aperture-synthesis imaging}
\label{sec:imaging}

\begin{figure}
  \centering
  \includegraphics[angle=0,scale=0.55]{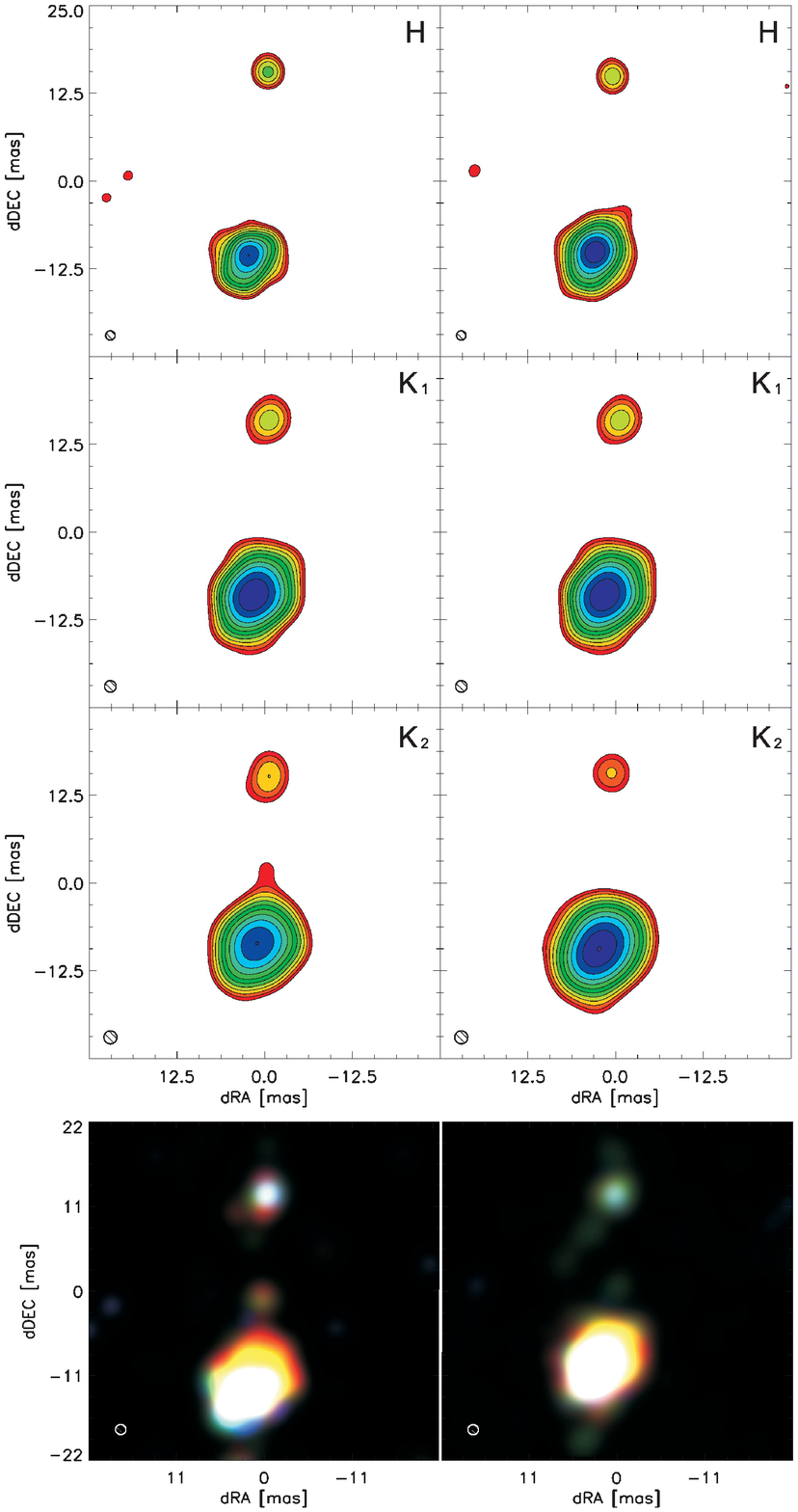}
  \caption{
    Aperture-synthesis images constructed without {\it (left column)}
    and with maximum entropy regularization {\it (right column)} from our
    VLTI/AMBER data for the $H$ {\it (top)}, $K_1$ {\it (2nd from top)}, 
    and $K_2$ {\it (3rd from top)} wavelength bins.
    The contours decrease from peak intensity by factors of $\sqrt{2}$.
    In the bottom row, the three wavelength bins are merged in a
    color composite (red: $K_2$, green: $K_1$, blue: $H$).
    All images were convolved to the formal resolution $\lambda/2B$,
    which is 1.3~mas ($H$), 1.6~mas ($K_1$), and 1.8~mas ($K_2$).
  }
  \label{fig:imaging}
\end{figure}

In order to derive the basic source structure of V921\,Sco,
we reconstructed model-independent aperture-synthesis images
from our AMBER data.
Since the object structure might change with wavelength,
we subdivide our data set in three wavelength bins, which we denote with
$H$ ($1.4 \leq \lambda < 1.9~\mu$m), 
$K_{1}$ ($1.9 \leq \lambda < 2.15~\mu$m), and
$K_{2}$ ($2.15 \leq \lambda < 2.5~\mu$m).
For each wavelength bin, we reconstruct a separate image (Fig.~\ref{fig:imaging}, {\it top})
and then combined the independent images in a color composite (Fig.~\ref{fig:imaging}, {\it bottom}).

For the image reconstruction, we employed the
building block mapping algorithm \citep{hof93},
which was already used in several of our earlier long-baseline
interferometric imaging projects \citep{kra07,kra09a,kra10}.
The presented images were obtained without (Fig.~\ref{fig:imaging}, {\it left})
and with (Fig.~\ref{fig:imaging}, {\it right}) 
regularisation function.
Image reconstruction with regularisation means the 
minimization of the cost function
\begin{equation}
  J[o_k({\bf x})] := Q[o_k({\bf x})] + \mu\cdot H[o_k({\bf x})],
\end{equation}
where $Q[o_k({\bf x})]$ describes the $\chi^2$ function of the 
measured bispectrum data $O^{(3)}({\bf f_u},{\bf f_v})$ and the 
bispectrum of the actual iterated image $o_k({\bf x})$. 
$H[o_k({\bf x})]$ is a regularisation term, 
and $\mu$ is a weighting factor called the Lagrange multiplicator.
For our reconstructions we used the maximum entropy
regularisation function
\begin{equation}
  H[o_k({\bf x})] := \int \left[o_k({\bf x})\cdot \log\left({\frac{o_k({\bf x})}{p({\bf x})}}\right) - o_k({\bf x}) + p({\bf x})\right] d{\bf x}.
\end{equation}
As prior function $p({\bf x})$ we used a smooth version of a 
building block reconstruction obtained without a 
regularisation function ($\mu = 0$).
The start image was a circumsymmetric Gaussian with a size obtained by 
fitting the measured visibilities.
Reconstructions for different Lagrange multiplicators ($\mu = 10^{-6}$ to $10^{-3}$) 
and for different reconstruction windows (radii: 60\,mas to 100\,mas) were obtained.
The presented images are the best fit reconstructions, i.e.\ those reconstructions 
with minimum reduced $\chi^2$ values of the squared visibilities and CPs.

Each of the reconstructed images clearly reveals a close companion,
which is located at a separation of $\sim 25$~mas
north of the primary star\footnote{In the Letter, we define the primary star (=V921 Sco A) as the star, which is associated with the majority of the near-infrared emitting circumstellar material.} ($\Theta \sim 353{\deg}$)
and which we denote in the following V921\,Sco~B.
In our images, V921\,Sco~B appears point-like, while the 
primary star is clearly associated with extended emission
that appears elongated along position angle (PA)
$\phi \sim 145{\deg}$.
In the images, it is also evident, that the contributions of V921\,Sco~B
to the total flux decrease with wavelength.

\subsection{Model fitting}
\label{sec:modeling}

In order to better characterize the geometry and physical conditions of
the disk-like structure in our image and to derive the relative astrometry 
of V921\,Sco~A-B, we fitted geometric models to our interferometric data.

In all models, the primary star is included as a point source, where the photospheric emission 
at a given wavelength is given by the flux ratio of a 
B0 ($T_{\rm eff}=14,000$~K, $g=4.04$) Kurucz model atmosphere 
\citep{kur70} to the measured total SED flux $F_{\rm tot}$, with contributions $F_{\rm A}/F_{\rm tot}$ 
ranging from 0.27 (at 1.5~$\mu$m), 0.13 (at 2.0~$\mu$m) to 0.07 (at 2.5~$\mu$m),
where $F_{\rm A}$ denotes the photospheric flux contribution of the primary star.
We adopt $A_V=4.8\pm0.2$, a stellar radius of $R_{\star} = 17.3\pm0.6~R_{\sun}$,
and a distance of $d=1150\pm150$~pc \citep{bor07}.
Given that the spectral type of the primary was determined in earlier studies
without knowledge of the companion star, we note that these spectral type estimates
are potentially biased, which could affect the photospheric flux of the primary star in our models.
However, the absolute level and wavelength-dependence of the photospheric emission in the near-infrared
wavelength regime depends only weakly on the precise spectral type and will therefore
not significantly affect our modeling results.

The companion star is described by five free parameters, namely the 
PA ($\Theta$) and separation ($\rho$) measured from the primary star, 
the angular extent of the circumstellar material
(parameterized as Gaussians with FWHM $\theta_{B}$),
and two parameters ($\frac{F_{\rm B}}{F_{\rm tot}}(\lambda_{\rm ref})$, $s$) 
to describe the photospheric flux contributions of V921\,Sco~B to the total flux.
The flux contributions might change with wavelength due to differences in the stellar effective temperatures, 
local extinction effects, or circumstellar emission.
Given the still relatively short wavelength coverage, we approximate these effects with the linear relation
$\frac{F_{\rm B}}{F_{\rm tot}}(\lambda) = \frac{F_{\rm B}}{F_{\rm tot}}(\lambda_{\rm ref}) + s \cdot (\lambda - \lambda_{\rm ref})$,
where we choose $\lambda_{\rm ref}:=2~\mu$m as arbitrary reference wavelength.

Free parameters of the circumprimary disk are the PA ($\phi$), 
angular extend along the major axis (Gaussian FWHM $\theta_{A}$),
and inclination angle ($i$, measured from the polar axis).
Due to temperature gradients in the circumstellar material,
it is possible that the source brightness distribution might change with wavelength.
To test this hypothesis, we performed our geometric fits both to the complete data set 
and to the aforementioned data subsets ($H$, $K_1$, $K_2$ wavelength bins).

In order to find the best fit, we employ a Levenberg-Marquardt least square fitting procedure
and minimize the likelihood-estimator $\chi_{r}^2 = \chi_{r,V}^2 + \chi_{r,\Phi}^2$, where 
$\chi_{r,V}^2$ and $\chi_{r,\Phi}^2$ are the reduced least square between the measured and model
visibilities and CPs, respectively (eqs.\ 1 and 2 in \citealt{kra09b}).
For our best-fit solution (Tab.~\ref{tab:modelfitting}), the
uncertainties on the individual parameters have been estimated 
using the bootstrapping technique. The fit provides a slightly better 
representation of the CP than the visibility-data, as indicated by 
the $\chi_{r,V}^2$ and $\chi_{r,\Phi}^2$-values.
Likely, this indicates that the circumprimary disk geometry is not well represented
by a Gaussian-brightness distribution and we will consider more physical
models in an upcoming study (Kraus et al., in prep.).

Given that the position of the companion might change notably between 2008 and 2009,
we first fitted only the 2008 data in order to characterize the properties of the 
stellar components and the circumstellar material.

\begin{deluxetable*}{cccccccccccccc}
\tabletypesize{\scriptsize}
\tablecolumns{14}
\tablewidth{0pc}
\tablecaption{Model-fitting results for the VLTI/AMBER continuum observations (Sect.~\ref{sec:modeling})\label{tab:modelfitting}}
\tablehead{
                         &                 & \multicolumn{5}{c}{Secondary star}                                        && \multicolumn{3}{c}{Circumprimary disk}                                          &                     &                       & \\
\cline{3-7} \cline{9-11}
  \colhead{Epoch}        & \colhead{Spectral} & \colhead{$\Theta$}        & \colhead{$\rho$}          & \colhead{$\theta_{\rm B}$}  & \colhead{$\frac{F_{\rm B}}{F_{\rm tot}}$}
                                                                                                                & \colhead{$s$}            && \colhead{$\theta_{\rm A}$} & \colhead{$i$}       &  \colhead{$\phi$}       &  \colhead{$\chi^{2}_{\rm r,V}$} & \colhead{$\chi^{2}_{\rm r,\Phi}$} & \colhead{$\chi^{2}_{\rm r}$} \\
                         & \colhead{band}            & \colhead{[\deg]}          & \colhead{[mas]}           & \colhead{[mas]}            & \colhead{(at $2~\mu\mathrm{m}$)} & \colhead{[$\mu$m$^{-1}$]} && \colhead{[mas]}           & \colhead{[\deg]}    & \colhead{[\deg]}       &                     &                       & \\
}
\startdata
2008                     & $H$              & 353.0          & 25.0            & $\leq 0.5$       & 0.074           & 0\tablenotemark{a}             && 6.12          & 51.1          & 143.2         & 2.46                & 1.40                  & 1.73\\
2008                     & $K_1$            & 353.4          & 25.0            & $\leq 0.5$       & 0.070           & 0\tablenotemark{a}             && 6.85          & 48.6          & 143.5         & 2.50                & 2.14                  & 2.25\\
2008                     & $K_2$            & 353.1         & 25.0             & $\leq 0.5$       & 0.064           & 0\tablenotemark{a}             && 7.55          & 48.5          & 149.0        & 3.91                & 1.97                  & 3.27\\
2008                     & all              & $353.8$       & $25.0$          & $\leq 0.2$       & $0.054$          & $-0.0056$        && $7.5$         & $50.3$        & $147.8$      & 4.30                & 3.09                   & 4.88\\
                         &                  & $\pm1.6$      & $\pm0.8$       &                   & $\pm0.018$       & $\pm0.019$       && $\pm0.2$      & $\pm1.9$      & $\pm4.3$     &                     &                        & \\
  \hline
2009                     & all              & $347.3$       & $25.5$         & $\leq 0.2$\tablenotemark{a}     & $0.054$\tablenotemark{a}         & $-0.0056$\tablenotemark{a}      && $7.5$\tablenotemark{a}       & 50.3\tablenotemark{a}        & 147.8\tablenotemark{a}     & 4.17            & 1.78                   & 3.39\\
                         &                  & $\pm1.0$      & $\pm1.2$       &                  &                   &                 &&                &                &               &                 &                        & \
\enddata
\tablenotetext{a}{In our fitting procedure, this parameter was kept fixed.}
\end{deluxetable*}

\subsection{Orbital motion measurement}
\label{sec:orbitalmotion}

Following the detailed characterization of the V921\,Sco system for epoch 2008
(Sect.~\ref{sec:modeling}), we then investigated whether we find evidence for
orbital motion between 2008 and 2009.
Compared to 2008, a significantly smaller number of AMBER observations was recorded
in 2009 (Tab.~\ref{tab:obslog}),
which lead us to fix the geometry of the circumprimary disk using the
best-fit parameters from epoch 2008 and treat only the two astrometric parameters 
($\Theta$, $\rho$) and the flux ratio ($F_{\rm B}/F_{\rm tot}$) as free fitting parameters.

The resulting best-fit parameters can be found in Tab.~\ref{tab:modelfitting} and
indicate that the position of the secondary has moved significantly in PA 
($\sim 7$\deg) during the covered $\sim 8$~months, while only marginal changes in separation 
($\rho \lesssim 0.5$~mas) or in the flux ratio could be detected over this time period.

\section{Interpretation}
\label{sec:interp}

\subsection{Characterization of the detected companion star}
\label{sec:interpcompanion}

Assuming a face-on circular orbit, we can estimate from the detected signs of orbital motion 
(Tab.~\ref{tab:modelfitting}) the period of the orbit to $P \sim 35$~yrs.
Of course, for a precise mass determination, long-term follow-up observations
will be necessary in order to derive the full dynamical orbit.

Both in our images and our model fits, the companion appears spatially
unresolved (Gaussian FWHM $<0.3$~mas, corresponding to $<0.35$~AU at 1.15~kpc), 
which suggests that V921\,Sco~B is not associated with circumstellar material.
Based on this result, we can compute the flux ratio of V921\,Sco~A and B for the 
$H$-band ($(F_{\rm B}/F_{\rm A})_{H}=0.83\pm0.15$) and 
$K$-band ($(F_{\rm B}/F_{\rm A})_{K}=1.18\pm0.12$).
These values suggest that V921\,Sco~B is of cooler temperature and 
later spectral type than the primary V921\,Sco~A, although the current
uncertainties are still too large for a quantitative spectral classification.

\subsection{Relation with large-scale structures}
\label{sec:interplargescale}

Our IMACS narrowband and broadband images 
(Fig.~\ref{fig:IMACS}) reveal a remarkably complex environment
around V921\,Sco.
The overall shape of the nebula is bipolar, 
where the south-western part appears much fainter and less extended,
maybe indicating that this part is facing away from the observer and therefore suffers from
a larger amount of obscuration from material in the ambient cloud.
Obscuration from fore-ground material might also be responsible for the 
dark filaments which appear in the south-eastern part of the nebula 
(shaded area in Fig.~\ref{fig:IMACS}{\it G}).

Remarkably, the symmetry axis of the bipolar nebula ($50\pm5^{\circ}$) appears well aligned
with the polar axis ($57.8\pm4.3^{\circ}$) of the AU-scale disk detected with interferometry,
suggesting that the nebula might have been shaped through outflow-activity from V921\,Sco.
From our kinematical modeling (Kraus et al., in prep.), we conclude that the
disk is notably inclined, with the north-eastern disk axis facing towards Earth.
Therefore, based on the measured disk inclination and orientation, one would expect the
north-eastern lobe of the bipolar nebula to appear brighter, as is observed.
This scenario provides a natural explanation for the large number
of arc- and cone-shaped structures which can be seen in our IMACS images (Fig.~\ref{fig:IMACS})
and that appear centered on V921\,Sco.
We have identified the most significant of these arc- and cone-shaped features in our deepest image 
($R$-band, Fig.~\ref{fig:IMACS}{\it E-F}).  
After confirming the features in images taken with other filters, 
we marked them in Fig.~\ref{fig:IMACS}{\it G}.

The distribution of the individual arc-fragments, in particular in the north-eastern lobe,
suggests that we might observe up to five layers of ejecta, which have been 
created during episodic events of extreme mass-loss.
In order to obtain a rough estimate for the period of the mass-loss events,
we measured the typical separation between the different layers ({4-5\arcsec})
and apply a projection factor in order to correct for the system inclination
angle of $50.3\pm1.9^{\circ}$ (Sect.~\ref{tab:modelfitting}).
Assuming the maximum expansion velocity of 1400~km\,s$^{-1}$ measured by \citet{bor09} and
a distance of 1.15~kpc, we find that a mass-loss period of $\sim 25$~yrs
would be required, which seems consistent with the period of the companion,
as estimated for a circular orbit (Sect.~\ref{sec:interpcompanion}).
Compared to the bow-shock structures observed around other high-mass YSOs \citep[e.g.][]{kra10}, 
the arcs around V921\,Sco have a clumpy and sometimes truncated structure,
possibly indicating a lower degree of collimation in the V921\,Sco outflow.

In a field of about one arcminute, we detect in our IMACS and FIRE images
at least 26 stellar sources. \citet{hab03} obtained spectra
for a subset of these sources (see identification in Fig.~\ref{fig:IMACS}{\it I}) 
and classified them as embedded low- and intermediate-mass YSOs.
Increasing the number density in the cluster, our observations strengthen
the argument that V921\,Sco is likely in a young evolutionary stage
and, as many massive YSOs, embedded in a star forming region of 
low- and intermediate-mass YSOs.

\subsection{Binary interaction and the B[e] phenomenon}
\label{sec:interpbinaryinteraction}

Our detection of a companion around V921\,Sco is interesting in the 
context of earlier suggestions that the B[e] phenomenon might be related to dynamical interaction 
in a close binary system, either as result of a recent stellar merger \citep[e.g.][]{pod06}
or through material which might have been ejected during phases of 
binary interaction \citep[e.g.][]{she00,mir07,kra10b}.
Unfortunately, only very few B[e]-stars have been studied so far 
with sufficient angular resolution to make definite statements about multiplicity.
One of the best-studied B[e] systems is HD\,87643,
where \citet{mil09} detected a close companion around this supergiant star.
Both the projected separation ($\sim 50$~AU) and orbital period ($\sim 50$~yrs) of
this system shows similarities with the characteristics
which we deduce for V921\,Sco ($\sim 29$~AU, $P \sim 35$~yrs).
Also, both systems are embedded in an extended nebulosity with
shell-like sub-structure, indicating episodic mass-loss.
Therefore, it is plausible that the mass-loss in both systems might be
triggered by dynamical interaction of the companions with the circumprimary disks.
The B[e]-star-characteristic permitted and forbidden line emission might then
originate in low density material which is periodically stripped away from the circumprimary disk.
This scenario works independently of the evolutionary status and physics of the circumprimary disks 
(accretion or excretion disks) and might explain the diversity of stellar systems 
(Herbig B[e], supergiant, symbiotic star, compact planetary nebulae), 
in which the B[e] phenomenon is observed.

Follow-up interferometric studies on V921\,Sco and HD\,87643 might confirm this scenario
for instance by measuring the orbital eccentricity and by imaging the expected star-disk 
interaction effects during periastron passage.

\section{Conclusions}
\label{sec:conclusions}

We have investigated the milliarcsecond-scale environment around 
the B[e] star V921\,Sco and summarize our findings as follows:

\begin{itemize}
\item In our model-independent interferometric imaging, we discover a close
  ($\sim 25.0$~mas) companion around the B-type star
  and detect signs of orbital motion ($\Delta\Theta \sim 7$\deg) between the 2008 and 2009 observations.
  The newly discovered companion V921\,Sco~B is apparently not associated with a
  circumstellar disk and of later spectral type than the primary, as indicated
  by the measured $H-K$ color.

\item Around the primary star, our images show a spatially extended
  disk-like structure, seen under an intermediate inclination angle of $50.3\pm1.9$\deg.

\item As about half of all B[e] stars \citep{mar08},
  V921\,Sco is associated with an extended nebula.
  In the case of V921\,Sco, the nebula has an intriguing bipolar morphology,
  where the symmetry axis ($50\pm5^{\circ}$) coincides with the polar axis
  of the AU-scale disk resolved by our interferometric observations ($57.8\pm4.3^{\circ}$).
  In our narrowband- and broadband-filter images (Fig.~\ref{fig:IMACS}), 
  we detect complex, partially arc-shaped sub-structures,
  which might have been shaped by episodic mass-loss events,
  possibly triggered by the periastron passage of the 
  newly-discovered close companion.

\item Our findings add new support to the hypothesis that the B[e] phenomenon 
  might be a consequence of interaction effects in close multiple systems,
  where the forbidden line-emitting material is ejected during close companion encounters
  and then episodically repeated during the periastron passage.
\end{itemize}

\acknowledgments

We thank Wen-Hsin Hsu for sharing some Magellan/IMACS observing time.
This work was done under contract with the California
Institute of Technology (Caltech), funded by NASA through the 
Sagan Fellowship Program (S.K.\ is a Sagan Fellow).

{\it Facilities:} \facility{VLTI}, \facility{Magellan}.

\bibliographystyle{apj}

\end{document}